\documentclass[review,12pt]{elsarticle}

\usepackage{amssymb}

\usepackage{lineno}

\usepackage{ulem}

\usepackage{graphicx}
\graphicspath{{./Figures/}}
\DeclareGraphicsExtensions{.pdf,.jpeg,.png,.tif}
\usepackage{color}
\usepackage{subcaption}
\usepackage{multirow}
\usepackage{multicol}   
\usepackage[cmex10]{amsmath}
\usepackage{bm}

\usepackage{array}

\usepackage[colorinlistoftodos,prependcaption,textsize=tiny]{todonotes} 

\journal{Journal of Marine Systems}

\begin{document}

\begin{frontmatter}

%

\title{Ensemble model aggregation using a computationally lightweight machine-learning model to forecast ocean waves}

\author[label1]{Fearghal O'Donncha \corref{cor1}}
\ead{feardonn@ie.ibm.com}
\cortext[cor1]{
Corresponding author \newline}
\author[label2]{Yushan Zhang}
\author[label1]{Bei Chen}
\author[label3]{Scott C. James}
\address[label1]{IBM Research - Ireland, Dublin}
\address[label2]{{University of Notre Dame}}
\address[label3]{Baylor University}



%


\begin{abstract}
This study investigated an approach to improve the accuracy of computationally lightweight surrogate models by updating forecasts based on historical accuracy relative to sparse observation data. Using a lightweight, ocean-wave forecasting model, we created a large number of model ensembles, with perturbed inputs, for a two-year study period. Forecasts were aggregated using a machine-learning algorithm that combined forecasts from multiple, independent models into a single ``best-estimate'' prediction of the true state. The framework was applied to a case-study site in Monterey Bay, California. A~learning-aggregation technique used historical observations and model forecasts to calculate a weight for each ensemble member. Weighted ensemble predictions were compared to measured wave conditions to evaluate performance against present state-of-the-art. Finally, we discussed how this framework, which integrates ensemble aggregations and surrogate models, can be used to improve forecasting systems and further enable scientific process studies.
\end{abstract}


%

\end{frontmatter}




\section{Introduction}
In recent decades, science has made significant advances in  enabling machines to understand language and process images for applications such as facial recognition, image classification, text translation, and autonomous driving. More generally, machine learning applied to other scientific domains, particularly the geosciences, are in a nascent stage. Some examples include applying artificial neural networks to simulate earthquake-cycle activity \citep{devries2017enabling}, accelerating Eulerian fluid simulations using convolutional neural networks \citep{tompson2016accelerating}, up-scaling air pollution forecasting using deep learning \citep{haehnel2018scaling}, and research to combine physics-based rules to guide machine-learning models of temperature profiles in a lake \citep{karpatne2017physics}.

A~computationally lightweight emulator (a surrogate for more complex systems) has a number of advantages, such as near real-time simulations over long time scales (e.\,g.,~multi-year studies of ocean conditions); the computational expedience of surrogate models is also important for systems affected by uncertainties in initial and boundary conditions such as risk-assessment studies that require iterative scenario modelling \citep{Fang2006,Simpson2008,Marrel2015}. A~wide range of analyses across different disciplines can benefit from such reduced computational complexity \citep{razavi2012review}. Examples where uncertainty is a major factor include Bayesian model calibration \citep{Bliznyuk2008,Khu2003}, global sensitivity analyses \citep{Blatman2010,Ratto2007,Marrel2015}, and Monte Carlo-based uncertainty or reliability analyses \citep{Shrestha2009,Schultz2006,Borgonovo2012}. 

There is a growing trend in the geosciences to combine deterministic and probabilistic tools to provide stakeholders with most likely forecasts together with confidence bounds on alternative outcomes \citep{raftery2016use}. This is particularly true in oceanography where large uncertainties exist related to fundamental aspects such as the physics of wind input, dissipation, and nonlinear interactions \citep{odonncha2015characterizing}. Here, we combine ensemble-forecasting and machine-learning techniques to: (1)~investigate uncertainty from an ensemble modelling system with perturbed inputs, (2)~leverage the advantages of computationally lightweight surrogate models, and (3)~generate a forecast that is better than the best individual model prediction.

The authors recently developed and demonstrated a machine-learning surrogate model for a physics-based ocean-wave model \citep{James2018-ML}. The model generated a nonlinear mapping of inputs (i.\,e.~wave height, period, and direction boundary conditions, spatially variable ocean currents, and wind speeds) to computed outputs (spatially variable significant wave height, $H_\mathrm{s}$, and characteristic wave period, $T$). The machine-learning model yielded enormous speedup (\textgreater five-thousand-fold) in computational time while maintaining accuracy that was well within the confidence bounds of the physics-based model.

Ensemble forecasts of wave conditions are typically generated from statistical perturbations of wave-height boundary data, ocean-current input data, wind forcing (particularly for global models), model physics, discretisation, and parameterisation schemes \citep{chen2006ensemble}. The fundamental objective of ensemble forecasting is to investigate inherent uncertainty to provide more accurate information about future states. This process facilitates transition from single, deterministic forecasting with optimistic assumptions on the fidelity of model inputs, to a multiple, probabilistic forecasting approach that realistically considers errors and uncertainties in the model forcing data and fundamental governing equations. Ensemble aggregation techniques can extend from simple arithmetic averages of all models to  machine-learning approaches that admit aggregate ensemble predictions based on weighted summation \citep{mallet2009ozone}. The learning-aggregation technique makes use of historical observations and model forecasts to produce a weight for each model. A~linear, convex (i.\,e.,~where weights are constrained so they sum to unity) combination of model forecasts is performed with these weights to generate the best model forecast. 

This study focused on an ensemble forecasting approach applied at a case-study site in Monterey Bay, California. Ensembles were created based on a careful analysis of model sensitivity to input data at three buoy locations. Model aggregation considered three approaches: (1)~na\"{i}ve model aggregation, (2)~ridge-regression forecasting, and (3)~forecasting using the exponentiated-gradient method. The paper presents a comprehensive framework to develop and aggregate ensemble model elements cognisant of the inherent uncertainties of inputs.

Our objective is to leverage the advantages of computationally lightweight surrogate models and non-invasive, ensemble aggregation techniques to provide the best-estimate of the system state. The contributions of the paper are as follows:
\begin{itemize}
    \item Framework that combines computationally lightweight surrogates and ensemble modelling to improve predictive skill.
    \item Ensemble generation implementation to fully explore the solution space based on case-study conditions and sensitivity to model boundary conditions.
    \item Analysis of data-driven ensemble aggregation techniques and the improvement in predictive skill compared to deterministic approaches.
\end{itemize}


\section{Methodology}
\label{sec:Methods}
In this section, we briefly describe the study site; development of the surrogate ocean wave model is described in detail elsewhere \cite{James2018-ML}. Section \ref{subsec:ensemble} describes the creation of the model ensemble elements whereby the surrogate model is run multiple times with perturbed inputs based on an analysis of  system dynamics. Finally, we describe the aggregation techniques adopted that considered two different methods to compute weights for each ensemble model element based on historical agreement with observations.

\subsection{Study site}
\label{sec:site}
Monterey Bay is a wide and partly deep ($>$1000 m) embayment in central California, broadly open to the ocean. It is an important economic and ecological location home to a federally protected national marine sanctuary (montereybay.noaa.gov) and a number of state marine sanctuaries \cite{chao2009development}. It is a well-studied site with several National Oceanic and Atmospheric Administration National Data Buoy Center (NOAA NDBC) buoys providing measurements of wave and wind characteristics dating as far back as 1987. The relative richness of observational data and well-studied dynamics makes Monterey Bay an ideal location to explore data-driven approaches.

\begin{figure}[h!]
\centering
\includegraphics[width=\textwidth]{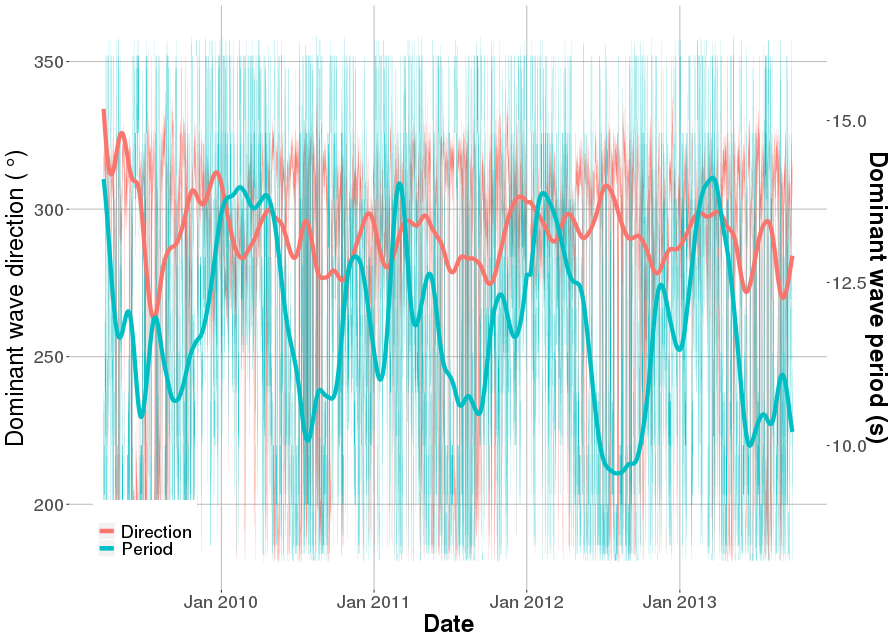}
\caption{Mean wave period (blue curves) and direction (red curves) at the outermost buoy (Buoy 46042 in Figure \ref{fig:Bathymetry}). The $y$ axis on the left-hand side represents direction while the right quantifies wave period. To better visualise primary trends, low-pass-filtered values are also shown (heavy curves)}
\label{fig:per_dir}
\end{figure} 

The oceanography of Monterey Bay is closely tied to processes of the California Current, a 1,200\,km broad and 300\,m deep surface current that transports water of subarctic origin southward along the North American coast at 15 to 30\,cm/s \cite{Broenkow1996}. Within the coastal regime, sea surface flow undergoes a seasonal reversal. During the late autumn and winter the direction is primarily to the north, while flow to the south dominates during the spring and summer driven by the intensification of northwesterly winds. These oceanographic trends influence wave conditions, with greater wave-heights during winter, and higher numbers of short-period locally generated waves during winter than during summer \cite{xu1999local}. Figure~\ref{fig:per_dir} plots measurements of dominant wave period and direction from a site in the outer bay (Buoy 46042 q.\,v. Figure \ref{fig:Bathymetry}) over a five-year interval. Observed waves were a combination of long-period (10--20\,s) swells and short-period (8--10\,s) locally generated wind waves with period varying between eight and 20 seconds. The low-pass filtered values demonstrated these seasonal trends with the raw data exhibiting the fluctuating characteristics of oceanic waves. 
A~prediction framework for ocean waves aims to capture both longer-term seasonal trends and small-scale fluctuations of these processes.

\subsection{Machine-learning Surrogate Model}
\label{sec:mldesc}
One of the challenges of applying machine learning to the geosciences is the enormous volumes of data required to train the model notwithstanding the amount of sensor data that can be available, which are often spatially sparse and intermittent. However, when developing a machine-learning surrogate for a physics-based model, there is the luxury of being able to run the model as many times as necessary to develop a sufficient data set for training. The Simulating WAves Nearshore (SWAN) model is the industry-standard wave-modelling tool developed at the Delft University of Technology that computes wave fields in coastal waters forced by wave conditions on the domain boundaries, ocean currents, and winds \citep{swan_tech_man}. 

To learn features of wave conditions, a supervised machine-learning model based on a multi-layer perceptron (MLP) network was used to compute significant wave heights, $H_\mathrm{s}$. During training, where the network was presented with examples of the computation it was learning (i.\,e.,~SWAN model runs), an optimisation problem was solved until the output of the network's last layer consistently approximated the training data set (in this case, the $H_\mathrm{s}$ field). Within a supervised-learning framework, inputs and outputs of the model were provided and the model learned optimised weights and biases that replicate the nonlinear relationship between inputs and outputs.

\begin{figure}[b!]
\centering
\includegraphics[width=1.0\textwidth]{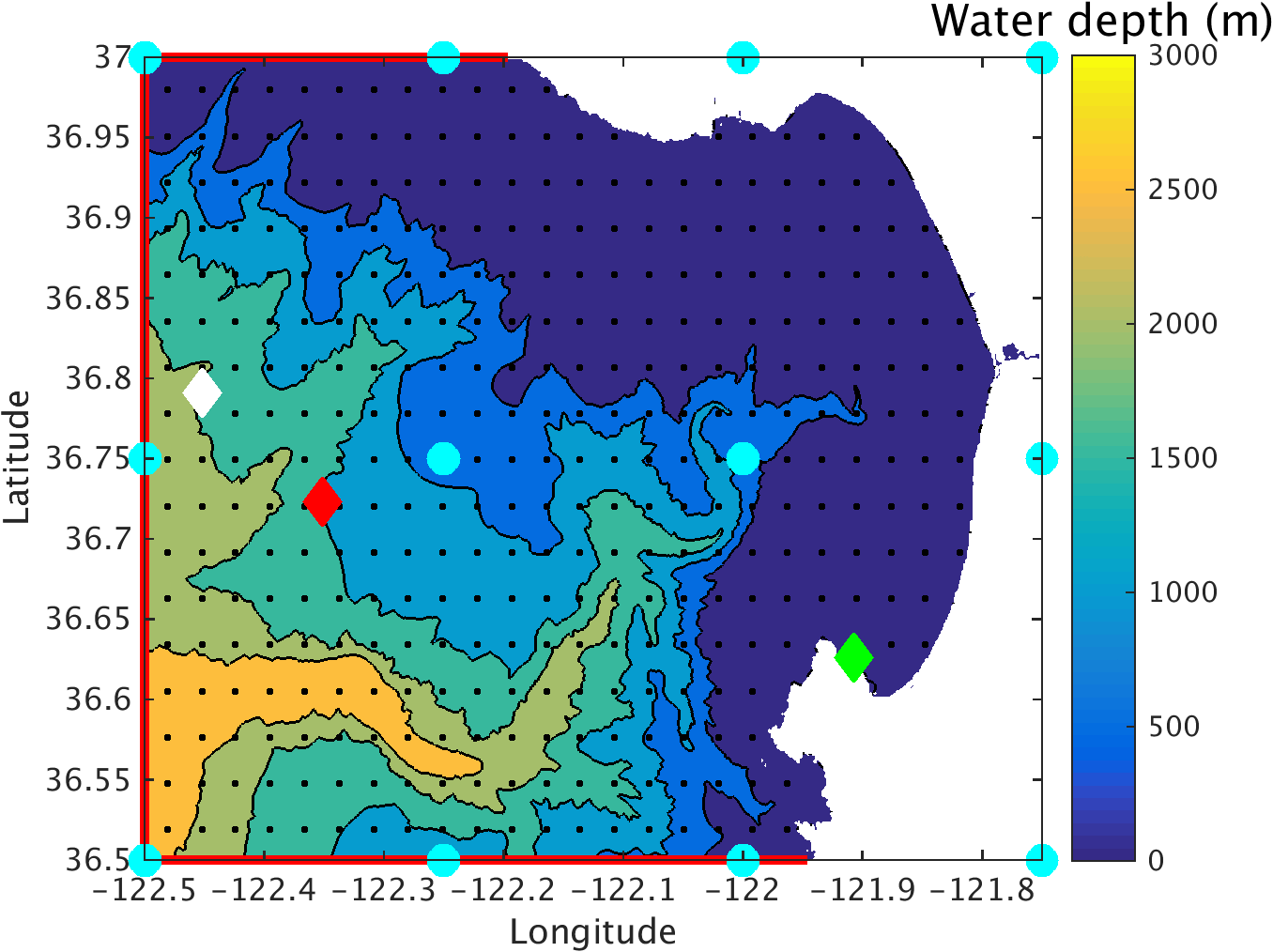} \\
\caption{SWAN model domain with color indicating bathymetric depth. The three buoys used to verify the model are indicated with the symbols where the white diamond is Buoy~46042, the red diamond is Buoy~46114, and the green diamond is Buoy~46240. Wave characteristics, $H_\mathrm{s}$, $T$, and direction, from nearby Buoy 46042 were specified along ocean boundaries indicated in red, TWC winds were specified at the 12 turquoise circles, and currents were specified at the 357 nodes of the ROMS model indicated with black dots.}
\label{fig:Bathymetry}
\end{figure}

The MLP \citep{goodfellow2016deep} model was trained at the Monterey Bay study site. Figure~\ref{fig:Bathymetry} illustrates the modelling domain \mbox{($64\times 54$-km\textsuperscript{2}} discretised across $71 \times 48$ computational elements providing a horizontal resolution of $0.01^\circ$ each approximately equal to $900 \times 1,000$\,m$^2$) for the SWAN model originally developed and validated by \citet{chang2016numerical}. NOAA NDBC Buoys 46042 (white), 46114 (red), and 46240 (green) provided measurements of wave conditions together with other ocean and meteorological data reported every 30 to 60\,minutes. Inputs to the SWAN model comprised boundary-condition wave data extracted from Buoy~46042 (Figure \ref{fig:Bathymetry}), ocean-current data from a Regional Ocean modelling System (ROMS) hydrodynamic model of Monterey Bay \citep{patterson2012addressing}, and historical wind data from The Weather Company (TWC). These were assembled into machine-learning input vectors, $\mathbf x$, with outputs corresponding to the SWAN-simulated $H_\mathrm{s}$ field, $\mathbf y$. Design matrices were developed by completing 11,078 SWAN model runs dating back to the archived extent of ROMS currents nowcasts (from April 1\textsuperscript{st}, 2013 to June 30\textsuperscript{th}, 2017) and stacking $\mathbf x$ and $\mathbf y$ into design matrices $\mathbf X$ and $\mathbf Y$, respectively.

Because the goal of this effort was to develop a machine-learning framework to act as a surrogate for the SWAN model, the nonlinear function mapping inputs to the best representation of outputs, $\hat{{\mathbf{y}}}$, was sought:
\begin{equation}
\label{eqn:1}
	  	g\left({{\mathbf x}};{{\mathbf{\Theta}}}\right)=\hat{{\mathbf{y}}}.
\end{equation}

A~sufficiently trained machine-learning model yields a mapping matrix, $\mathbf \Theta$, that acts as a surrogate for the SWAN model. This facilitated sidestepping of the SWAN model by replacing the solution to its partial differential equation of wave energy density with the data-driven machine-learning model composed of the vector-matrix operations encapsulated in (\ref{eqn:1}). 

MLP regression was used to reproduce the SWAN-generated $H_\mathrm{s}$ field, ${\hat{\mathbf{y}}}$. Fundamental performance was assessed by comparing against the spatially variable SWAN $H_\mathrm{s}$ predictions, ${\mathbf{y}}$. The ${{\mathbf{X}}}$ and ${{\mathbf{Y}}}$ data were randomly shuffled into two groups to form the training dataset composed of 90\% of the 11,078 rows of data with the testing dataset the remaining 10\%. Mapping matrix ${{\mathbf{\Theta}}}$ was calculated using the training dataset and then applied to the testing dataset and the RMSE between test data, $\mathbf{{y}}$, and its machine-learning estimate, $\mathbf{{\hat{y}}}$, was calculated. 

The machine-learning model demonstrated excellent performance against the SWAN model with  RMSE $\approx 9$\,cm, notably less than the confidence bounds of the model (about 40 to 50\,cm \citep{bidlot2002intercomparison}). Further, the computational time to make forecasts was reduced by a factor of 1/5,000\textsuperscript{th}. Comparison against measured buoy data revealed the main limitation of surrogate modelling -- a surrogate model will never outperform the model it emulates and the most well-designed surrogate can only aspire to match the accuracy of the more expensive model. 


\subsection{Integrating Forecasts with Machine-learning Aggregation}
\label{sec:Methods_ML}
\subsubsection{Creation of Model Ensembles}
\label{subsec:ensemble}
As discussed previously, ensemble forecasting typically focuses on multiple simulations where anything from physical parametrisations, numerical discretisation, or input data are perturbed. Machine learning models do not readily admit adjustments to model parameters (indeed, that contravenes the data-driven philosophy of the approach); instead, ensemble construction focused on perturbation of model inputs. Inputs to the ML surrogate model consisted of lateral boundary information on wave height, period and direction, and spatially variable ocean currents and wind speeds, which were prescribed at hourly intervals over the simulation period. Previous studies investigating the sensitivity of the Monterey Bay SWAN model to perturbed inputs of wind forcing (extracted from the NOAA Global Ensemble Forecast System or GEFS) demonstrated low sensitivity to perturbation of wind and ocean current inputs \citep{donncha2017-ensemble}. As a result of the limited spatial scale of the Monterey Bay model domain ($\approx3,500$\,km\textsuperscript{2}), perturbing wind input data based on outputs from NOAA GEFS forecasts yielded changes in wave height of less than 0.5\,cm. Considering this lack of sensitivity, and to simplify the ensemble-generation process, only lateral-boundary wave data were perturbed.

\begin{figure}[t!]
\centering
        \includegraphics[width=\textwidth]{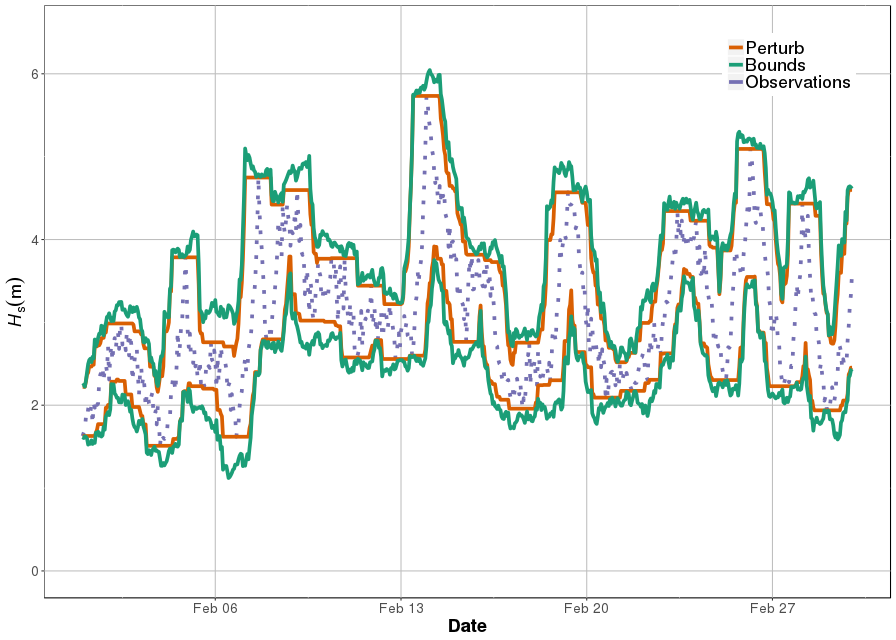}
\caption{Time series of $H_\mathrm{s}$ from which perturbed values were selected for boundary-data specification. The blue points are measured data, the red curves span the daily maximum/minimum, and the green curves include Gaussian perturbations to incorporate variable system dynamics.}
\label{fig:bounds}
\end{figure}

As described in Section~\ref{sec:mldesc}, lateral boundary data to the model were prescribed based on wave measurements extracted from Buoy 46042 (Figure \ref{fig:Bathymetry}). This served as the deterministic, or benchmark, model for the study (best individual model). 
Creation of perturbed wave boundary conditions emanated from these data by: (1)~defining the upper and lower thresholds and (2)~quantifying system dynamics based on the spread around the wave time-series data, which were superimposed on the above thresholds. Upper and lower thresholds for $H_\mathrm{s}$
were defined as the maximum and minimum daily values (red curves in Figure \ref{fig:bounds}). To this, a random perturbation was added to represent the statistical dynamics of the system. The perturbation was estimated from a random Gaussian process model \citep{rasmussen2004gaussian} with distribution of mean zero and spread computed from the 48-hour rolling window standard deviation of the observations. 

Final upper and lower thresholds (green curve in Figure \ref{fig:bounds}), were generated from combination of these perturbations and boundary wave height data for model ensembles were sampled from within these limits. Sixteen model ensembles were created by sampling boundary wave height data using a standard Latin hypercube sampling (LHS) technique; a statistical method for generating a near-random sample of parameter values from a distribution \citep{mckay2000comparison}. LHS proceeds by dividing a distribution or range into $N$ (in this case 16), equally probable intervals (in this case, the range at each point in time is bounded by the green curves in Figure \ref{fig:bounds}) and a random sample was selected from within each interval.  This ensured adequate coverage of a distribution where the tails are important -- a pertinent consideration for wave forecasting where accurately predicting maximum wave heights may be more important than capturing ambient conditions.

\begin{figure}[h!]
\centering
\includegraphics[width=\textwidth]{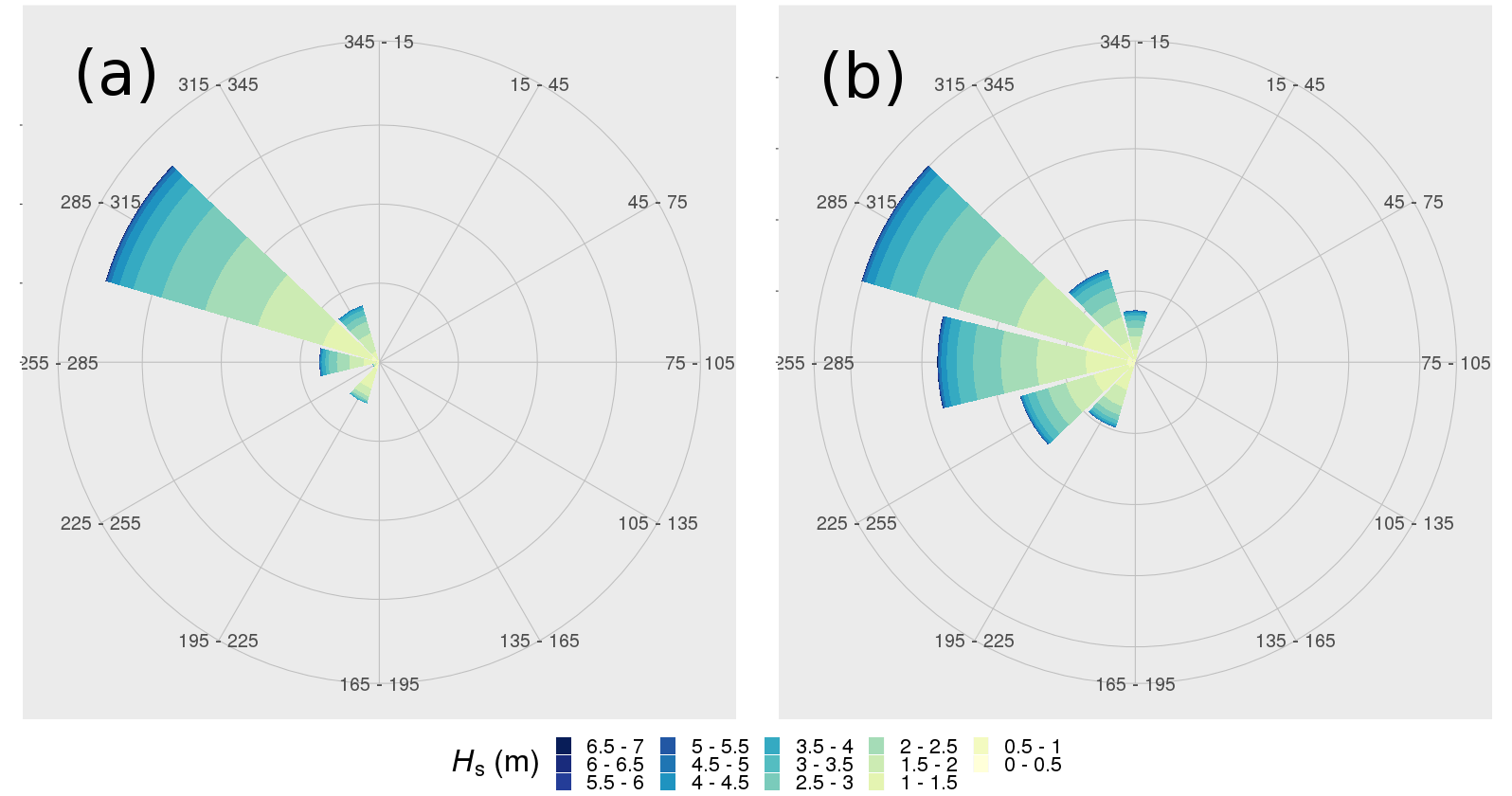}
\caption{(a)~Northwest-prevailing wave directions during the study period. (b)~Boundary-condition wave direction perturbations were randomly selected from  the four bins. Directions are presented in the meteorological convention.}
\label{fig:windrose}
\end{figure} 

Further analysis of model forecasts indicated that the ensembles with perturbed wave-height forcing did not fully capture dynamics at the near-shore Buoy 46240 (analysed in more detail by \citet{donncha2017-ensemble}). In particular, the model failed to capture wave conditions developed when swells were from the southeast. To ensure the ensemble spread fully captured the dynamical space, wave directions were perturbed in a similar manner to wave heights, namely:
\begin{itemize}
    \item Upper and lower thresholds were created from the daily maximum and minimum of measured wave directions.
    \item A random component was added to represent sensor uncertainty and system dynamics. As for wave height, the random component was sampled from a normal distribution of mean zero and spread computed from the standard deviation of the measured wave direction.
\end{itemize}
These produced upper and lower limits on wave directions at each point in time from which wave directions were sampled using the LHS technique. Six ensemble elements with perturbed directional inputs were created. A~total of 119 ensemble elements were generated -- 16 with perturbed wave heights, six with perturbed directions, and one each with unperturbed height and direction.

\begin{figure}[t!]
\includegraphics[width=\textwidth]{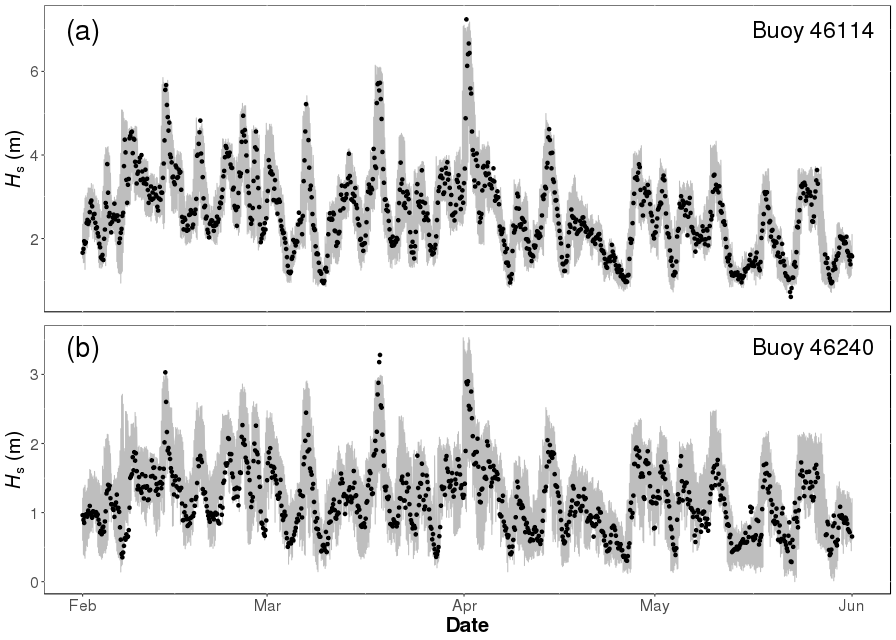}
\caption{Time-series representation of generated ensembles (116~individual elements) upon perturbing $H_\mathrm{s}$ and $D$. Curves represent individual forecasts for each ensemble member and the black circles denote observations from Buoy  (a) 46114  and (b) 46240.}
\label{fig:genensemble}
\end{figure}

Classical works on ensemble forecasting demonstrated that the ensemble mean should give a better forecast than a single deterministic forecast as long as the ensemble represents the uncertainty present in the forecast \citep{epstein1969stochastic, leith1974theoretical}. It is therefore critical that the ensemble spread encapsulate observed dynamics. Ensemble generation methods generally focus on uncertainties in: (1)~initial conditions and (2)~forcing data (uncertainty of model equations and parametrisation are also often considered). Ensemble methods with perturbed initial conditions are ubiquitous in meteorology and these focus on quantifying the fastest growing errors with techniques such as the breeder method and optimal perturbation analysis, common in data-assimilation implementations \citep{turner2008ensemble}. Specification of lateral boundary conditions for ensemble models is usually either from ensemble predictions on larger domains (such as the NOAA GFS) or from  perturbation around a deterministic estimate or observation \citep{torn2006boundary}. \citet{evensen2003ensemble} described the required characteristics of an ensemble set of models as: including sufficient spread to capture pertinent dynamics, and that each model or ensemble element be linearly independent.

For an ensemble aggregation study, the ensemble spread and relationship between individual model elements is a key consideration.
Figure~\ref{fig:genensemble} presents the generated ensemble predictions against measured data for a representative six-month period at the outer-bay and near-shore buoy locations.
The ensemble spread captured by the models was clearly sufficient to encapsulate observed dynamics, particularly at the near-shore buoy that proved more sensitive to perturbed forcing. 
It is worth noting that one of the primary focus of the ensemble generation was ensuring that system dynamics were fully encapsulated by the ensemble spread -- this allowed a pragmatic approach whereby the ensemble generation was ``hand crafted'' to some degree to analyse and understand the different combinations of wave height and directions that best captured observed spread at each buoy. This approach was assisted by the low computational cost of the surrogate model (compared to the full SWAN model) where different ensemble predictions could be readily computed and analysed -- making a two year prediction with all of the 119 ensemble elements took less than 18 minutes compared to 290 days that would be required to run the full SWAN model (on equivalent compute resources). Characteristics of individual ensemble elements were observed to be a complex interplay of prescribed wave height and direction together with the nonlinear dynamics of the MLP model (recall that other forcing such as winds and ocean currents were kept constant across different ensemble elements). The objective of the ensemble aggregation can be considered as computing a set of weights to reduce the ensemble set to a unique forecast that minimises differences between forecasts and (highly dynamic and variable) observations at multiple buoy locations.

\subsubsection{Model aggregation}
The underlying assumption of ensemble modelling is that each member contains some information pertinent to the true state of the system. The interplay between models is expected to vary in both space and time; i.\,e.,~member models performed better at different points in space and time depending upon ambient conditions, individual model forcing, and other physical interactions. 
Ordinary least squares (OLS) is a widely used approach that estimates model predictions that minimise the sum of squares of residuals (against observations). The variance of OLS estimates is high as it preferentially weights the models that yield the best predictive skill at each point in time (conceptually overfitting to the data). The focus in this study was to develop an aggregation framework that incorporated historic model performance while also accounting for seasonal or shorter-term wave condition dynamics. In effect, the aggregation balances historic performance and short-scale dynamics to aggregate the ensemble predictions into a single ``best-estimate'' forecast.

The first technique investigated was a ridge-regression (RR) prediction algorithm. In this approach, a weight vector was computed for each time update that minimised the difference between past predictions and observations. The weight vector for each time was \citep{mallet2009ozone}:
\begin{equation}
\mathbf{u}_t =  \arg\min_{\mathbf{u} \in R^n} \left[\lambda \vert \vert \mathbf{u} \vert \vert_{2}^{2}
+ \sum\limits_{t^\prime=1}^{t-1}  \sum\limits_{s \in S_{t^\prime}} 
\left( \mathbf{u} \cdot \mathbf{x}_{t^\prime}^s - y_{t^\prime}^s \right)^2  \right],
\label{eqn:2}
\end{equation}
where $\mathbf{x}^s_{t^\prime}$ is a vector of dimension $\mathcal N$ (number of ensembles equal to 116) containing the prediction from each ensemble member at time $t^\prime$ (for each station $s$), ${y}^s_{t^\prime}$ represents observations at time $t^\prime$ and station $s$, $\mathbf{u}_t$ is vector of weights computed for each ensemble member at time $t$, $S_{t^\prime}$ represents the number of observation stations for which data are available at each time, and $\lambda$ serves as a regularisation constant used to keep the magnitude of $\mathbf{u}_t$ small and to reduce variation between consecutive vectors.  \citet{mallet2009ozone} described this regularisation (penalty) function, which is typically selected in a {{bespoke}} manner for each study to balance contributions from the most recent model-observation datasets and historical data. As $\lambda$ tends toward zero, the regression tends toward an OLS solution.  Conceptually, the objective was to assign weights to each ensemble member that minimised the MSE across observation stations (Buoys 46114 and 46240). Training the weights vector progressed on all data to $t-1$ whereupon predictions were made for the next time step, $t$, based on the most recent ensemble predictions. 

The computed weights were then used to make a forecast, $\hat{x}^s_t$, for each station, $s$, at time $t$ as:
\begin{equation}
\hat{x}^s_t =   \mathbf{u}_t \cdot \mathbf{x}_t^s =   \sum\limits_{m=1}^{\mathcal N} {u_{m,t}x_{m,t}^s},
\label{eqn:3}
\end{equation}
where $\mathbf{x}_t^s$ is each member of the ensemble prediction at station $s$, and $\mathbf{u}_t$ is the weight vector applied to each prediction. 

The second technique explored here was an exponentiated-gradient (EG) algorithm for linear predictors \citep{kivinen1997exponentiated}. The EG algorithm also has a weight vector, $\mathbf{u}_t$, used to predict $\hat{x}^s_t = \mathbf{u}_t \cdot \mathbf{x}^s_t$. The updated weight for each ensemble member $x_{m,t^\prime}$ was \citep{kivinen1997exponentiated}:
\begin{equation}
u_{m,t} = \frac{  \sum\limits_{t^\prime=1}^{t-1} r_{m,t^\prime}u_{m,t^\prime}}{\sum\limits_{j=1}^{\mathcal N}  \sum\limits_{t^\prime=1}^{t-1} r_{j,t^\prime}u_{j,t^\prime}},
\label{eqn:4}
\end{equation}
for all $m=1,\ldots,\mathcal N$,  and: 
\begin{equation*}
r_{m,t^\prime} = \mathrm{exp} \left[ \sum\limits_{s \in S_{t^\prime}}  {-2\mu ( \mathbf{u}_{t^\prime} \cdot \mathbf{x}^s_{t^\prime} - y^s_{t^\prime})x^s_{m,t^\prime}} \right],
\label{eqn:5}
\end{equation*}
where $\mu$ is the learning rate. 

Weights computed with the EG approach were normalised by the sum of all weights as expressed by (\ref{eqn:4}). This constrains the weights to a convex combination as opposed to the unconstrained weights admitted by RR (i.\,e.,~where weights could take any values that minimised the loss function). A~potential advantage of EG-type approaches over RR is this constraint on weights, which limits rapid fluctuations. As opposed to unconstrained weights, convex-combination weight vectors may better extend to other regions of the model domain away from where observations are available \citep{mallet2009ozone}. EG predictions will always fall within the envelope of the ensemble predictions, which avoids unrealistic model forecasting. 

A~further aspect of the weight computation was selection of the historical window length. A~na\"ive approach uses all available historical data while more sophisticated implementations acknowledge that performance in the recent past can be more indicative of predictive skill. Amending (\ref{eqn:2}) for RR aggregation to incorporate user-specified window lengths, $t_\mathrm{w}$, yielded:
\begin{equation}
\mathbf{u}_t =  \arg\min_{\mathbf{u} \in R^n} \left[\lambda \vert \vert \mathbf{u} \vert \vert_{2}^{2}
+ \sum\limits_{t^\prime=t-t_\mathrm{w}}^{t-1}  \sum\limits_{s \in S_{t^\prime}} 
\left( \mathbf{u} \cdot \mathbf{x}_{t^\prime}^s - y_{t^\prime}^s \right)^2  \right],
\label{eqn:6}
\end{equation}
which differs from (\ref{eqn:2}) only in the starting index, $t-t_\mathrm{w}$. Similar to selection of the learning rate, $\mu$, and the regularisation constant, $\lambda$, a cross-validation approach was adopted to identify the optimum $t_\mathrm{w}$. We expand on the importance of judicious selection of these parameters and the challenges of providing general recommendations on appropriate choices in Section~\ref{sec:results}.

\section{Results and Discussion}
\label{sec:results}
Analysis of the results and performance of the aggregation technique focused on the two-year period, 2012--2014. Model aggregation consisted of an on-line forecasting tool using past data to update weights. Weights were initialised as zero vector or with value $1/\mathcal{N}$ (i.\,e.,~assigning equal weight to each prediction) for the RR and EG forecasters, respectively, and their values updated at each model forecast time based on a minimisation of the difference between forecast and observation.  
A number of prediction periods were considered to evaluate the performance of the approach in improving one-, two- and three-day dynamic forecasts. 

\begin{figure}
\centering
    \begin{subfigure}[b]{0.49\textwidth}
        \centering
        \includegraphics[width=\textwidth]{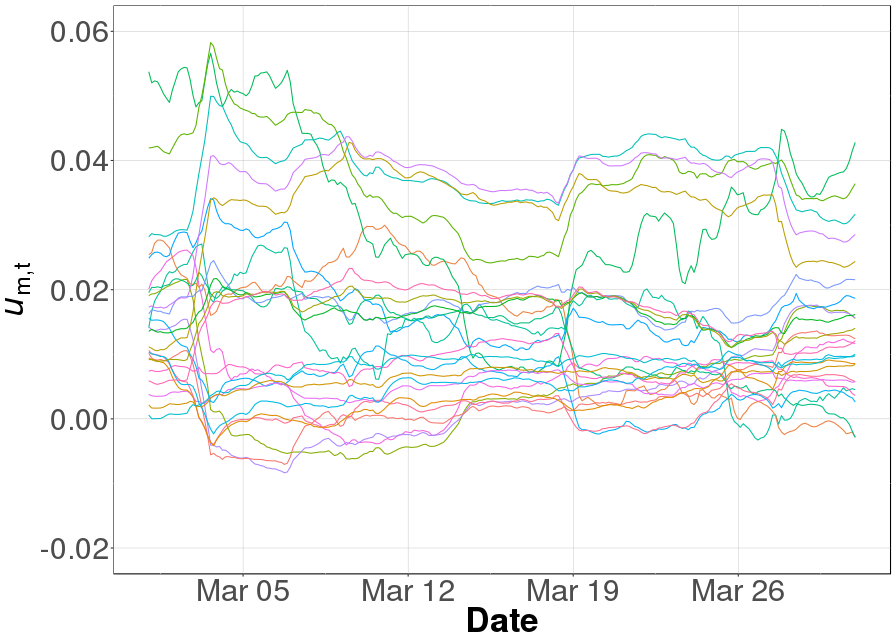}
    \end{subfigure} 
    \begin{subfigure}[b]{0.49\textwidth}
        \centering
\includegraphics[width=\textwidth]{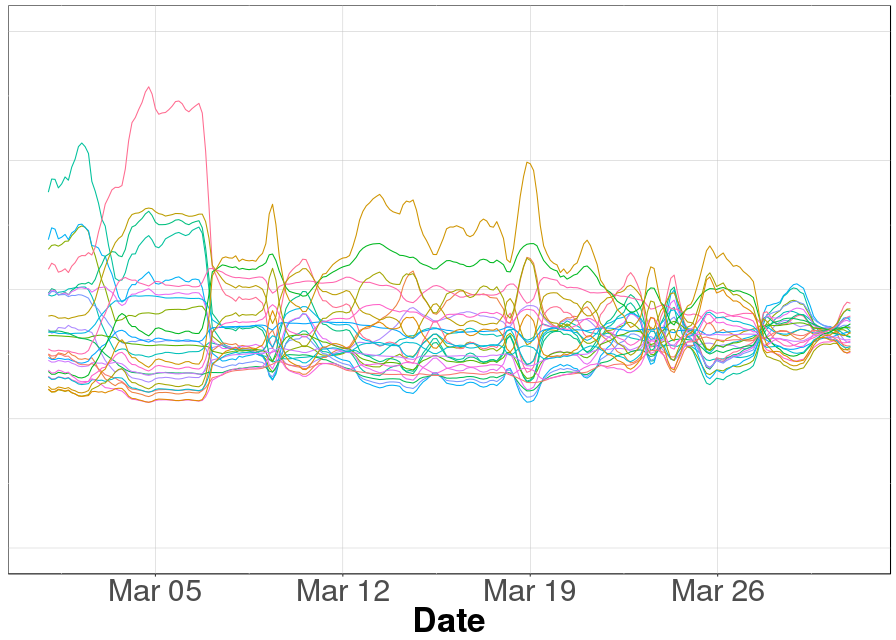}
    \end{subfigure}%
\caption{Weights computed for a representative subset (25) of the 116-member model ensemble over a one-month period. LHS represents RR weights while RHS represents those from EG. Each curve represents the weight, $u_{m,t}$,  attached to an individual model and shows its evolution over time.}
\label{fig:weights}
\end{figure}

Figure~\ref{fig:weights} presents a selection of model weights computed with the RR and EG aggregation methods for a representative one-month period. Most models contributed to the aggregated forecast with the majority of weights having strongly non-zero values. Further, the dynamics and variations of the weighting aggregation over time are apparent with larger-magnitude weights corresponding to periods with larger spreads in model forecasts (and consequently higher uncertainties from the ensemble-prediction perspective).  During periods of large model spread, minimisation of model-observation differences was facilitated by applying large weights to models that performed well and low weights to models that performed poorly. 
The time evolution and variations of weights were sensitive to a number of factors, namely, observation variance, regularisation coefficients, window length, and model prediction accuracy. Figure~\ref{fig:weights} demonstrates that EG weights tended to reduce in magnitude over the period presented (note that EG weights are constrained to sum to one). This is largely due to a combination of observation variance and the constrained nature of the weights for EG.

A~key consideration in wave forecasting is the temporal dynamics of the system. The fundamental basis of weighted model aggregation is that there is a certain relationship between successive forecasts and observations; i.\,e.,~there is a likelihood that the ensemble element that performed best at time $t$ will be the model that performs best at time $t+1$. By updating weights based on the difference between the latest observations and forecast, it ensured that the best-performing model was assigned the highest weight. 

\begin{figure}[t!]
\includegraphics[width=\textwidth]{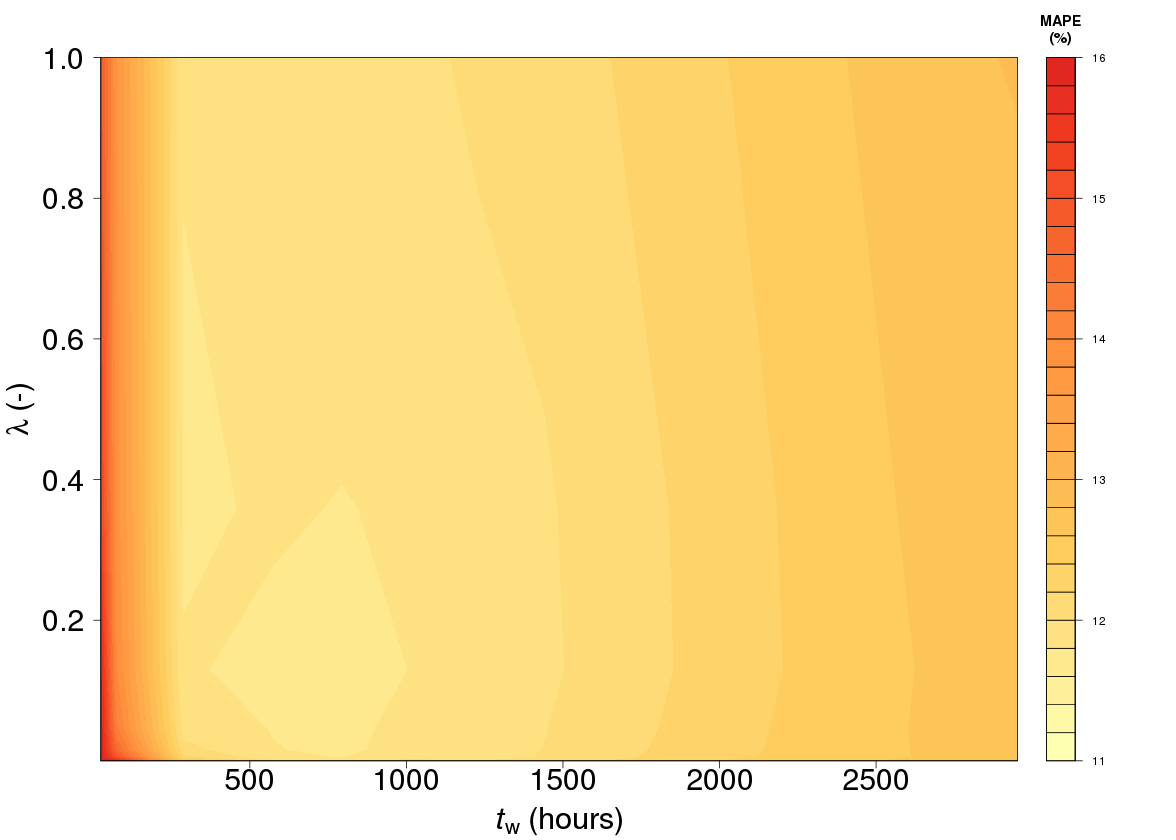}
\caption{Contour plot of MAPE computed from a 24-hour RR aggregation forecast as a function of  window size, $t_\mathrm{w}$, and regularisation constant  $\lambda$ for the two-year study period.}
\label{fig:mapewindowsize}
\end{figure} 

This ``follow-the-leader'' type forecasting system works best if the quantity being modeled is relatively stationary with pronounced historic influence -- ocean waves are highly dynamic with significant temporal variation. Hence, a key requirement of the approach was to select the parameterisation that best exploited historical performance while maintaining the ability to adjust to rapidly changing wave measurements. A~series of cross-validation experiments were used to select the most appropriate value for the regularisation constant or learning rate for RR and EG, respectively, and also to select the optimal $t_\mathrm{w}$. Namely, for both RR and EG, 500 experiments were conducted that varied $\lambda$, $\mu$, and  $t_\mathrm{w}$. To guide the selection process, minimum values were specified to avoid overfitting. For both $\lambda$ and $\mu$, a minimum value of $1.5\times10^{-10}$ was specified, while a minimum value of 24 hours was specified for $t_\mathrm{w}$.

For each cross-validation experiment, a 24-hour forecast was produced (creating a total of 730 forecasts for the two-year period), and the MAPE was computed against observations. 
Figure~\ref{fig:mapewindowsize} presents MAPE for RR as a function of  $\lambda$, and  $t_\mathrm{w}$.
Results demonstrated the need to balance short-term dynamics and model skill. As  $\lambda$ and  $t_\mathrm{w}$ tended toward zero, the aggregation reduces to a weight generation that minimised residuals between the model and observations (in effect, fitting the prediction to the data). This does not provide appreciable predictive skill as noise in the system can dominate (i.\,e.,~a model that happened to perform well at a given point in time tended to be weighted heavily and a model that had good long-term predictive skill could be  excessively punished for short-timescale deviations) and dynamic forecasting skill is sacrificed. For short  $t_\mathrm{w}$ ($<$ 48 hours), results demonstrated high MAPE as the computed weights were not representative of system dynamics. Predictive skill was not as sensitive to the specification of regularisation constant with a general trend of increased MAPE for $\lambda>0.6$. Results demonstrated that $t_\mathrm{w}=524$\,hours and $\lambda=0.18$  yielded the lowest MAPE.

More generally, appropriate specification of $\lambda$ and  $t_\mathrm{w}$ was required to balance the importance of historical performance and short-scale dynamics. In cases with limited sensor data (learning conditions across the bay from two data points), it is important to avoid overfitting. This is particularly relevant in this study where both buoy locations have distinct characteristics (outer-bay and near-shore dynamics). Hence, the cross-validation approach was combined with a judicious selection of parameters that leaned more heavily toward increased dependence on historic effects. Namely, the largest $\lambda$ and  $t_\mathrm{w}$ values that produced low MAPEs and that preceded a noticeable increase in MAPE were selected. For this case, $t_\mathrm{w}=928$\,hours and $\lambda=0.34$. This can be interpreted as an implementation of the ``elbow method'' commonly used in clustering approaches that aims to optimally balances error and computational complexity (of increased data size) \citep{bholowalia2014ebk}.

From these cross-validation studies, hyperparameters $t_\mathrm{w}$, $\lambda$, and $\mu$ that minimised MAPE were specified and forecasts were created from the aggregated model predictions. Results were compared against two benchmark forecasts:
\begin{itemize}
\item Best individual forecast defined as the model with unperturbed inputs.
\item Ensemble average consisting of the average of all ensemble forecasts (equivalent to applying a weight of $1/\mathcal{N}$ to each element).
\end{itemize}

\begin{figure}[h!]
\centering
    \begin{subfigure}[b]{0.7\textwidth}
        \centering
        \includegraphics[width=\textwidth]{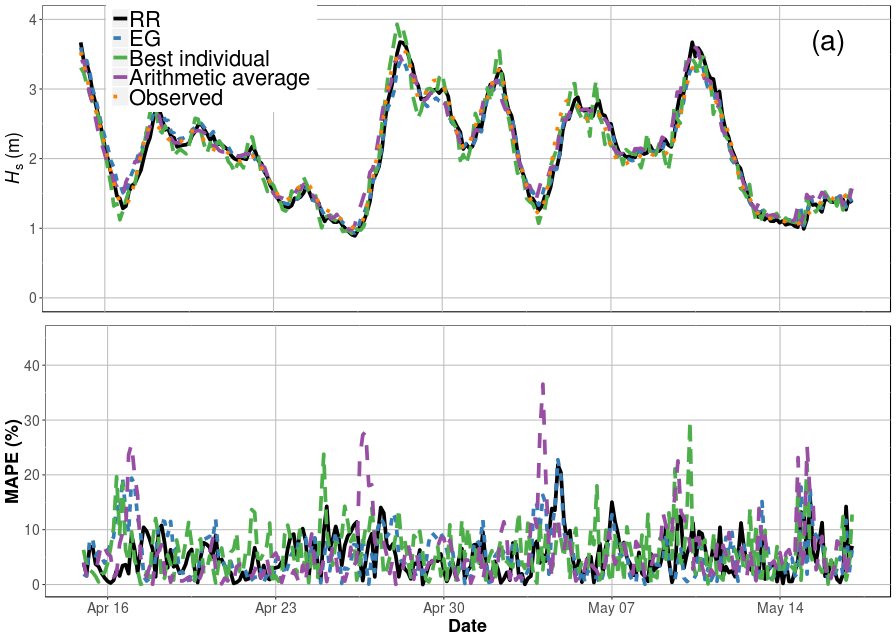}
    \end{subfigure} 
    \begin{subfigure}[b]{0.7\textwidth}
        \centering
\includegraphics[width=\textwidth]{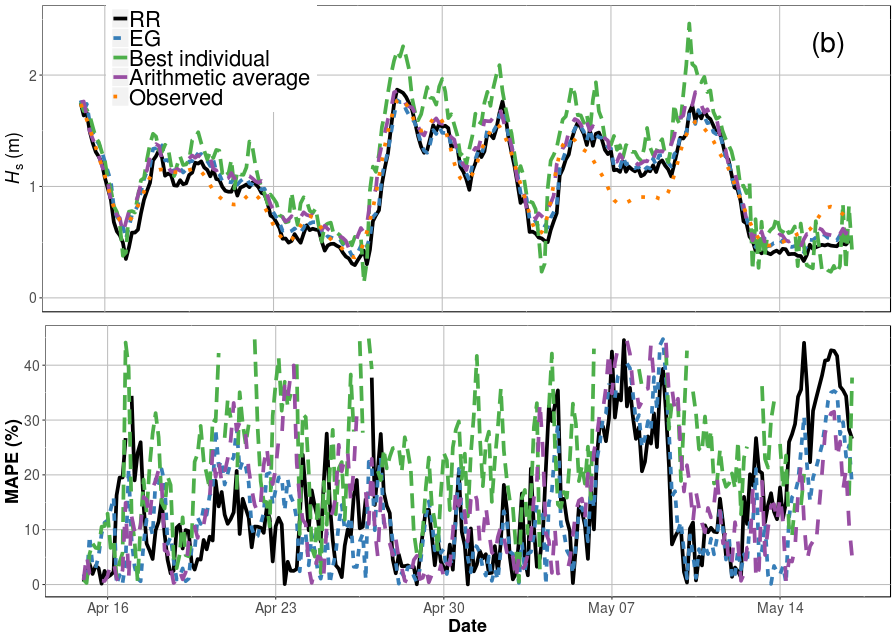}
    \end{subfigure}%
\caption{24-hour predictions of $H_\mathrm{s}$ (curves) plotted against observations (orange dots) at Buoy (a)~46114 and (b)~46240. The black curve denotes forecasts aggregated by RR. The dashed blue curve represents EG aggregation. The dashed green curve represents the best element from the 119-member ensemble, while the purple curve is the arithmetic average across all model forecasts. The top figure presents predictions plotted against observations, while the bottom figure presents MAPE for each prediction.}
\label{fig:modelComp}
\end{figure}

Figure~\ref{fig:modelComp} compares 24 hour forecasts at both sensor locations (Buoy 46240 and 46114 in Figure~\ref{fig:Bathymetry}) to observations. Results demonstrated that an aggregation that took into account historical model performance through weighted averaging outperformed more na\"ive approaches. MAPE was reduced from 22.2\% for a deterministic model with unperturbed inputs (green curve) to 11.7\% for the RR aggregation (black curve) and 12.1\% for the EG method (blue curve).  RR outperformed EG largely due to the unconstrained weights, which can more quickly react to rapid changes in a volatile system. This reduced errors, but may yield overfitting; the constrained nature of the EG weights may be more robust and applied farther from observation stations with more confidence than the unconstrained RR \citep{mallet2009ozone}. 

\begin{table}
      \caption{MAPE (\%) for each model scenario at each buoy, and averaged across both buoys. MAPEs are presented for three  prediction windows, a 24-, 48-, and 72-hour predictions, respectively (second row). The model scenarios presented are best individual ensemble model element (BI), arithmetic average of all ensemble elements (AA), and ridge regression (RR) and exponentiated gradient (EG) aggregation.}
\tiny
  \centering
  \begin{tabular}{|l|c|c|c||c|c|c||c|c|c||c|c|c|}
    \hline
    \textbf{Buoy} & \multicolumn{3}{c|}{\textbf{ BI}} & \multicolumn{3}{c|}{\textbf{ AA}} & \multicolumn{3}{c|}{\textbf{ RR}} & \multicolumn{3}{c|}{\textbf{EG}} \\
    \cline{2-13}
    & \textbf{24} & \textbf{48} & \textbf{72} & \textbf{24} & \textbf{48} & \textbf{72} & \textbf{24} & \textbf{48} & \textbf{72} & \textbf{24} & \textbf{48} & \textbf{72} \\
    \hline
    \textbf{46114}    &  7.1   & 7.1   &7.1  
                      & 7.6   &  7.6   & 7.6
                      & 6.7   & 6.9   &  7.2 
                      & 7.1   & 7.3   & 8.0 \\
     \textbf{46240}   & 37.3  & 37.3  & 37.3
                      & 23.7  & 23.7  & 23.7  
                      & 16.8  & 17.2  & 16.7
                      & 17.2  & 17.5  & 18.1 \\
     \textbf{Average} & 22.2   & 22.2  & 22.2 
                      & 15.6  & 15.6   & 15.6 
                      & 11.7  & 12.1   & 12.0
                      & 12.1  & 12.4  & 13.0\\
     \hline
  \end{tabular}
  \renewcommand{\arraystretch}{1.2}
  \label{table:MAPE}
\end{table}

Table \ref{table:MAPE} lists the MAPE for each model implementation for different forecasting windows.
It is worth noting that the weight computation selected those that provided best agreement averaged across buoys data rather than either in isolation. Further, each buoy have distinct environmental condition, namely, open-ocean conditions (Buoy 46114) and sheltered, near-land conditions (Buoy 46240), as evident in Figure \ref{fig:modelComp}. The resultant aggregation  generated a best-estimate forecast of ocean conditions across the available buoy locations and hence may be more applicable far from the buoy locations (i.\,e.,~at arbitrary locations within the bay).
For all forecasting windows, ensemble aggregation significantly outperformed a single forecast with unperturbed inputs. A na\"{\i}ve aggregation based on a simple ensemble average reduced MAPE by 30\%, while RR and EG aggregation provided a 47\% and 45\% reduction, respectively (for a 24 hour forecasting window). Predictive skill was comparable across different forecasting windows (24, 48, and 72\,hours) indicating the robustness of the computed weights and the ability to produce dynamic forecasts.

Two important assumptions were made in this study:
\begin{itemize}
    \item The observations are accurate and do not contain any sensor uncertainty or error and
    \item The accuracy of the model is similar throughout the spatial domain .
\end{itemize}{}
These assumptions meant that the aggregation provided equal importance to each observation point. Practical implementations may incorporate further considerations such as the evolution of uncertainty in observations over time and geographical regions of greater importance for forecasting (e.\,g.,~one may prefer a forecasting system that is most accurate near-shore). There is a large body of research on quantifying observation uncertainty, particularly in the data-assimilation literature \citep{evensen2009data}. Maintaining the common assumption that sensor errors are uncorrelated in time, appropriate specification of regularisation and window length ($\lambda$, $\mu$, and $t_\mathrm{w}$) parameters tends to exclude high-frequency errors by weighting based on long-term model-observation agreement. Observation errors that include both bias and variance are more difficult to incorporate and requires a model or estimate of observation bias (that can be removed from the data or used to specify confidence metrics).

Further, analysis of model performance at both buoy locations demonstrated much higher errors at the near-shore location due to more complicated physics and nonlinear interactions. Improving the resolution of near-shore dynamics in third-generation wave models is an active area of research. It is plausible in a coastal wave model that one may be more interested in accurately forecasting near-shore processes than those in the open bay. With a sufficiently dense network of near-shore sensors, aggregation can be tuned to specific forecasting requirements.
The framework can be readily extended with a weighting or confidence metric on individual sensor or stations based on such considerations.

These results demonstrated the viability of combining data-driven ensemble aggregation techniques with lightweight surrogate-model approaches to improve predictive skill. The  nominal computational expense of the machine-learning forecasting model means there is no real practical limit to the number of ensemble elements that can be generated (generating a seven-day forecast takes a fraction of a second). Recall that training the machine learning model on simulated data from a physics-based wave model (SWAN) established an upper limit on accuracy equal to that of the SWAN model, i.\,e.,~the surrogate model was trained to reproduce SWAN-predicted ocean waves and not real-world data as is generally the case in machine learning. However, by combining the surrogate model with ensemble-aggregation approaches that compute individual model weights based on predictive skill, this shortcoming can be overcome.

\section{Conclusions}
\label{sec:Concls}
In this paper, we detailed the creation and generation of ensemble predictions using a lightweight, machine-learning model. Identified limitations of the surrogate model were addressed by developing ensemble-aggregation techniques to minimise MSE against available measured data. RR and EG aggregating algorithms computed deterministic forecasts from the ensemble that leveraged past observations and past performance of each ensemble model. Results demonstrated that the aggregating forecaster notably reduced error against observations and it represents a valuable framework to integrate sparse sensor data with lightweight, data-driven surrogate models.

One of the primary advantages of this approach is that it provides a non-invasive method to leverage data to improve forecasts. As the algorithm only acts on outputs from the model to compute weighted-sum predictions, it does not require any update or propagation of the model state as is necessary with traditional data-assimilation approaches (and extremely difficult to integrate with machine-learning-based models) making it relatively straightforward to implement. Further, the aggregation algorithms can be readily replaced with alternative local-minima approaches that better reflect the needs of a particular study (e.\,g.,~gradient-descent approaches, etc.).

A~promising aspect of the approach not discussed here is the potential of using this framework to investigate long-term processes like the multi-decadal geosciences study by \citet{arandia2018surrogate}. Computationally lightweight surrogate models provide opportunity to investigate system dynamics across long time scales much easier than would be possible with large-scale models. The accuracy of these surrogate models can be improved with ensemble aggregation methods as presented here.

Ensemble-based forecasting is a widely used technique that accounts for uncertainty inherent in numerical modelling studies. Leveraging multiple simulations that encompass a wide range of potential scenarios facilitates an expanded exploration of likely future conditions and provides probabilistic information on forecasts. Many decision processes however, require a single, deterministic forecast. This is typically done with some form of averaging across all ensemble members or selection of the best individual model (based on some metric). The approach presented here outlines a comprehensive technique that leverages information from past model performance and observations to aggregate ensemble elements into a single forecast. The non-invasive framework can be easily integrated into an on-line operational forecasting system. This can be readily extended to other models and in particular to combining and aggregating models with different levels of complexity and different fundamental physics (e.\,g.,~combining rule-based models with data-driven models or deterministic approaches with stochastic). Future work will consider aggregation approaches that integrate models of different complexity and confidence metrics to provide prior information to the aggregating technique and parameterisation. Approaches that also consider sensor uncertainty and bias can extend the applicability of the framework for operational forecasting.

\section*{Acknowledgements}
\noindent
This project has received funding from the European Union’s Horizon 2020
research and innovation programme as part of the RIA GAIN project under grant
agreement No.~773330.

\section*{References}

\end{document}